\documentclass[twocolumn,amsmath,amssymb,10pt,superscriptaddress,a4paper,letterpaper,fleqn]{revtex4-1}
\usepackage{amssymb}
\usepackage{epsfig}
\usepackage{graphicx}
\usepackage{dcolumn}
\usepackage{array}
\usepackage{bm}
\usepackage{fancyheadings}
\usepackage{longtable}
\usepackage{multirow}
\usepackage{float}
\usepackage{afterpage}
\usepackage{color}
\usepackage{textcomp}

\setlongtables
\usepackage[breaklinks=true,linkbordercolor={1 1 1}]{hyperref}

\def\etal{et al.\,}

\begin{document}
\setcounter{page}{1}


\title{
\qquad \\ \qquad \\ \qquad \\  \qquad \\  \qquad \\ \qquad \\
Characterization of a Be(p,xn) neutron source for fission yields measurements.}

\author{A. Mattera}
\author{P. Andersson}
\author{A. Hjalmarsson}
\author{M. Lantz}
\author{S. Pomp}
\email[Corresponding author: ]{stephan.pomp@physics.uu.se}
\author{V. Rakopoulos}
\author{A. Solders}
\author{J. Valldor-Bl\"ucher}
\affiliation{Uppsala University, Uppsala, Sweden}

\author{D. Gorelov}
\author{H. Penttil\"a}
\author{S. Rinta-Antila}
\affiliation{University of Jyv\"askyl\"a, Jyv\"askyl\"a, Finland}

\author{A.V. Prokofiev}
\author{E. Passoth}
\affiliation{The Svedberg Laboratory, , Uppsala University, Uppsala, Sweden} 

\author{R. Bedogni}
\author{A. Gentile}
\author{D. Bortot}
\author{A. Esposito}
\affiliation{INFN-LNF Laboratori Nazionali di Frascati, Frascati (Rome), Italy}

\author{M.V. Introini}
\author{A. Pola}
\affiliation{Politecnico di Milano, Milano, Italy.}
\date{\today}

\begin{abstract}
We report on measurements performed at The Svedberg Laboratory (TSL) to characterize a proton-neutron converter for independent fission yield studies at the IGISOL-JYFLTRAP facility (Jyv\"askyl\"a, Finland). A 30~MeV proton beam impinged on a 5~mm water-cooled Beryllium target. Two independent experimental techniques have been used to measure the neutron spectrum: a Time of Flight (TOF) system used to estimate the high-energy contribution, and a Bonner Sphere Spectrometer able to provide precise results from thermal energies up to 20~MeV. An overlap between the energy regions covered by the two systems will permit a cross-check of the results from the different techniques. In this paper, the measurement and analysis techniques will be presented together with some preliminary results.
\end{abstract}
\maketitle

\lhead{}
\chead{}
\rhead{A. Mattera \textit{et al.}}
\lfoot{}
\rfoot{}
\renewcommand{\footrulewidth}{0.4pt}


\section{INTRODUCTION}
High-quality measurements of independent fission yields from key actinides are desirable at various neutron energies in view of the possible use of fast neutrons in generation-IV reactors and of innovative cycles for the handling of spent nuclear fuel. At the IGISOL-JYFLTRAP facility (Jyv\"askyl\"a, Finland) a new high intensity MCC30/15 cyclotron with proton beams up to 100~$\mu$A in current and 30~MeV in energy is now available, and a proton-neutron converter has been designed to provide suitable fields to study neutron induced fission yields of different actinides. The study of the fission products will be performed using the JYFLTRAP apparatus, which has been used for measurements of proton-induced independent fissions yields \cite{penttila2010epj}.

Different materials and geometries for the proton-neutron converter have been investigated, the preferred design being a 5~mm-thick water-cooled Beryllium target. The target is thinner than the stopping length of 30~MeV protons, allowing them to stop in the subsequent layer of cooling water. In this way, the cooling requirements and the hydrogen build-up in the metal, which could both represent major problems at high currents, are significantly reduced. The neutron yield has been studied with Monte Carlo calculations using MCNPX \cite{mcnp} and Fluka \cite{fluka1,fluka2} simulation codes \cite{andreas2013}, but given the large energy dependence of the fission cross-section, a good knowledge of the energy spectrum is required and hence, a direct measurement is desired.

The present work describes measurements performed at The Svedberg Laboratory (TSL), Uppsala (Sweden) to characterize the energy spectrum from a mock-up of the proton-neutron converter. 
\section{EXPERIMENTAL SETUP}
The available energy at IGISOL-JYFLTRAP was reproduced by degrading the 37.3~MeV proton beam of TSL down to 30.3~MeV, by means of a 1~mm thick aluminium tile.

The degraded proton beam was shaped by two sets of graphite collimators: a first x-y collimator 10~mm wide was used to reduce the size of the beam and a cylindrical collimator with an opening 15~mm in diameter, positioned in close proximity of the target assembly, was used to avoid any neutron production in the aluminium target holder.

Two independent experimental techniques were used to measure the neutron spectrum: a Time of Flight (TOF) system was used to estimate the high-energy contribution; and a Bonner Sphere Spectrometer (BSS), to measure neutrons energy from the thermal region up to 20~MeV. 

The BSS was composed of 9 spheres with radii between 2 and 12 inches. Details about this system and the results will be reported elsewhere \cite{bedogni2013}.

The TOF setup consisted of a 3.3~l BC-501 liquid scintillator from the NORDBALL array \cite{arnell1991NIMA} placed at different distances (1.19, 2.00 and 4.82~m) from the source, at an angle of 10 degrees with respect to the beam-line. The distances were selected to obtain a good energy resolution over the whole measured energy interval (see section \ref{sec:enres}) and to minimize the wrap-around effect given by the time structure of the cyclotron, that has a period of 44.25~ns. The Pulse Shape Discrimination (PSD) capabilities of the liquid scintillator helped reducing the gamma background.
Benchmarking for the TOF measurement system was provided with two simpler version of the neutron source: the same proton beam was made to impact on both a thin (1~mm) and a full-stop (6~mm) Beryllium target.

\begin{figure}[htb]
 \includegraphics[width = 0.49\textwidth]{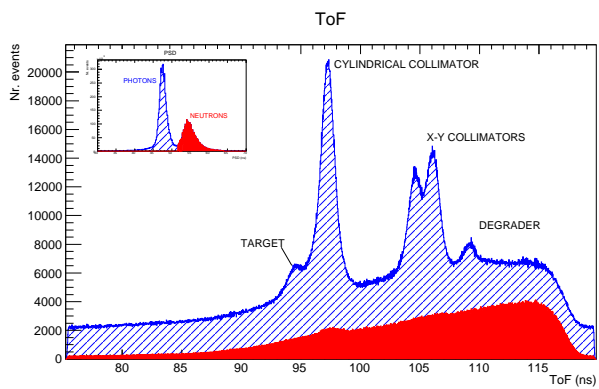}
\caption{Raw TOF spectrum. The PSD spectrum in the inset shows the separation between neutrons (solid red) and photons (hatched blue). The peaks in the photon spectrum correspond to production in the beam-line elements. The reduction of the gamma background is evident in the TOF spectrum.}
\label{fig:PSD}
\end{figure}

Two different Data AcQuisition (DAQ) systems were used simultaneously during the experiment. In both systems, the trigger coincided with an event above threshold detected in the scintillator and the stop signal came from the radio-frequency (RF) pulse from the cyclotron; the latter signal is delayed with respect to the time protons hit the target:
$$T_{\mathrm stop} = T_0 + \Delta t $$
where $T_0$ is the instant when protons reach the Be target and $\Delta t$ is the delay. It was possible to reconstruct the delay $\Delta t$ from the peaks corresponding to the gamma flash produced in objects along the beam-line (Fig. \ref{fig:PSD}).
The first analogue system provided online PSD with a dedicated NIM module \cite{arnell1991NIMA}. The information was stored along with the TOF, digitized by a TDC with a 25~ps sampling time, on an event-by-event basis. The energy threshold was varied between 3 values, that were estimated to correspond to about 2, 3 and 7~MeV neutrons.

The second acquisition system used a Multi Channel Analyser (SP Devices ADQ412 High-speed Digitizer) with which the whole pulse shape of each event was stored for off-line PSD. In this case, the threshold is adjustable off-line. The RF signal was also stored to extract the TOF information. The data presented in this paper come from the analogue DAQ: the analysis of the data collected with the Multi Channel Analyser is at an early stage.

\section{RESULTS}
\subsection{Measurement of the background}
\begin{figure}[htb]
 \includegraphics[width = 0.49\textwidth]{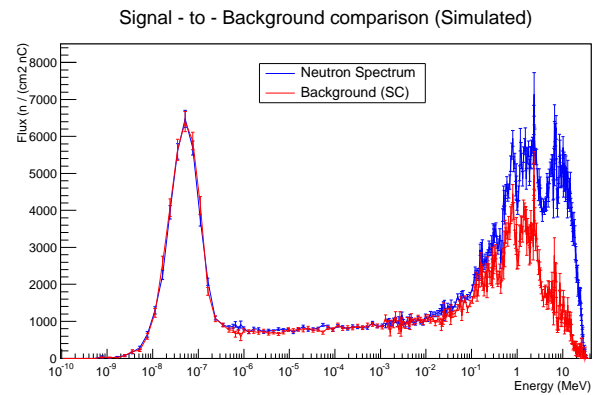}
\caption{Comparison of the FLUKA simulated neutron spectrum without (blue) and with (red) the shadow cones, to estimate the scattered neutrons background.}
\label{fig:SC-nSC-MC}
\end{figure}

Preliminary Monte Carlo simulations of the neutron transportation in the experimental hall showed that a significant background can be expected (Fig. \ref{fig:SC-nSC-MC}). This mainly originates from low energy neutrons produced in the collimators or other beam-line elements or from neutrons scattered in air and off the walls of the experimental area.

It was then important to estimate the contribution of all neutrons not coming from the Be(p,xn) source. With both the TOF system and the BSS, the background was estimated using the shadow-cones technique. 50~cm long shadow-cones (20~cm iron + 30~cm polyethylene) were manufactured in 5 different sizes to accommodate the dimension of the Bonner spheres and the scintillator and to ensure good shadowing conditions at all distances. 

\begin{figure}[htb]
  \includegraphics[width = 0.49\textwidth]{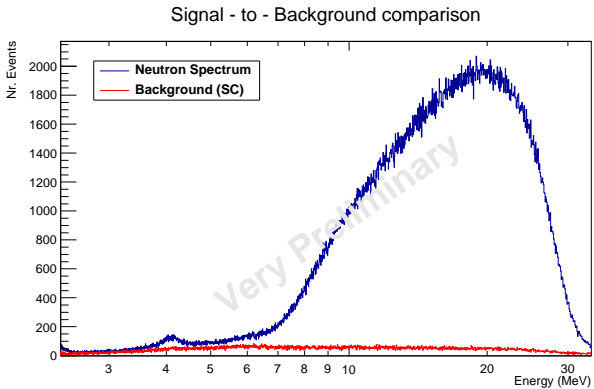}
\caption{Comparison of the measured neutron spectrum without (blue) and with (red) the shadow cones, to estimate the scattered neutrons background.}
\label{fig:SC-nSC}
\end{figure}

The first result for the TOF measurement is shown in Fig. \ref{fig:SC-nSC}. The threshold cut in the data presented here is such that the contribution from low energy neutrons, that are expected to dominate the background, is suppressed.

\subsection{Pulse Shape Discrimination}

The parameters of the PSD were optimized before the TSL run with a $^{252}$Cf n-$\gamma$ source and verified with the neutrons from the Be source. An example of the effect of the PSD is shown in Fig. \ref{fig:PSD}. Despite the very effective reduction of the continuum gamma background and in the lower gamma peaks, it can be noticed that the online PSD has some limitations in subtracting the intense gamma peak from the collimators and the target itself. This effect could not be explained by simple analysis of the analogue data and was observed for all the combination of distances and thresholds.

\subsection{Expected energy resolution}
\label{sec:enres}
In first approximation, the energy resolution of the TOF spectrometer depends on the accuracy with which time and distance can be estimated, according to the equation:
\begin{equation}
  \frac{\delta E }{E} = 2 \cdot \left[ \left( \frac{\delta L}{L} \right)^2 + \left( \frac{\delta T}{T} \right)^2 \right]^{\frac{1}{2}}
  \label{eqn:uncertainty}
   \end{equation} 
where L is the source-to-detector distance and T is the TOF.

For 20~MeV, equation \ref{eqn:uncertainty} gives an uncertainty of 4.2~\% at 4.8~m, that increases to 17~\% at 1.19~m; the uncertainty is dominated by $\delta L$, assumed equal to the half-thickness of the liquid scintillator $\approx 8~{\mathrm cm}$, while $\delta T \approx 1~{\mathrm ns}$ is dominated by the duration of the proton bunch, that was determined by a dedicated measurement with protons directly impinging on the detector. To improve the energy resolution, the position of interaction of neutrons should be known more accurately, but even in the most favourable scenario ($\delta L \approx 1~{\mathrm cm}$), the energy resolution cannot be expected to be better than 2.6~\% for the farthest distance.


\section{FUTURE WORK}
The limitations of the online-PSD will be investigated further by analysing the data collected with the MCA, where an off-line pulse shape analysis will be performed event-by-event. In addition to the PSD information and the TOF, the pulse height will help identifying the gamma background.

Monte Carlo simulations are being performed with different computer codes (MCNPX, FLUKA and GEANT4 \cite{geant4}) to be compared with the measured data. They will also be used to extract the response function of the detector in order to correct the spectra obtained with the detection efficiency at different energies.

\subsection*{ACKNOWLEDGEMENTS}


The  work was supported by the European Commission within the Seventh Framework Programme through Fission-2010-ERINDA (project no. 269499), by the Swedish Radiation Safety Authority (SSM), and by the Swedish Nuclear Fuel and Waste Management Co. (SKB).


\end{document}